\documentstyle[preprint,tighten,floats,citesort,aps]{revtex}

\newcommand{\beq}{\begin{equation}}
\newcommand{\eeq}{\end{equation}}
\newcommand{\mc}{\multicolumn}

\begin{document}
\draft

\title{The $\Delta I=1/2$ rule in non-mesonic weak decay of $\Lambda$-hypernuclei}
\author{W. M. Alberico$^{a,b}$ and G. Garbarino$^{b,c}$}
\address{$^a$ Dipartimento di Fisica Teorica, Universit\`a di Torino \\
$^b$ INFN, Sezione di Torino, 10125 Torino, ITALY\\
$^c$ Department of Theoretical Physics, Univ. of Oxford,
Oxford OX1 3NP - England}

\date{\today}
\maketitle

\begin{abstract}
By employing recent data on non-mesonic decay of $s$-shell $\Lambda$-hypernuclei
we study, within the phenomenological model of Block and Dalitz,
the validity of the $\Delta I=1/2$ rule in the 
$\Lambda N\to NN$ process. Due to the low experimental precision,
a possible violation of this rule can be neither proved nor excluded at present
with sufficient accuracy: a pure $\Delta I=1/2$ transition amplitude is
excluded at 40\% confidence level. 

\end{abstract}

The $\Delta I=1/2$ rule has been first established from the experimental
observation of the free $\Lambda$ decay rates into the two pionic
channels:
\begin{equation}
\label{lambdadec}
\Lambda \rightarrow
\begin{array}{ll}
 \pi^- p \qquad
&(\Gamma^{\mathrm free}_{\pi^-}/\Gamma^{\mathrm free}_{\Lambda}=0.639) \\
 \pi^0 n \qquad
&(\Gamma^{\mathrm free}_{\pi^0}/
 \Gamma^{\mathrm free}_{\Lambda}=0.358) ,
\end{array}
\end{equation}
the total lifetime being 
$\tau^{\rm free}_{\Lambda}\equiv \hbar/\Gamma^{\rm free}_{\Lambda}=
2.632\cdot 10^{-10}$ sec.
The experimental ratio of the above widths,
$(\Gamma^{\rm free}_{\pi^-}/\Gamma^{\rm free}_{\pi^0})_{\rm Exp}\simeq 1.78$,
as well as  the $\Lambda$ polarization observables  are suggestive of a
change in isospin of $\Delta I=1/2$, which also holds true for other
non--leptonic strangeness changing processes, like the decay of the $\Sigma$
hyperon and pionic kaon decays.

Actually this rule is slightly violated in the $\Lambda$ decay, since
(taking the same phase space for both channels and neglecting
the final state interactions) it predicts
$\Gamma^{\rm free}_{\pi^-}/\Gamma^{\rm free}_{\pi^0}=2$, a value
which is slightly larger than the experimental one. Nevertheless, the
ratio $A_{1/2}/A_{3/2}$ between the $\Delta I=1/2$ and the $\Delta I=3/2$
transition amplitudes for the decay $\Lambda \to \pi N$ is very large
(of the order of $30$). 

This isospin rule is based on experimental observations, but its
dynamical origin is not yet understood on theoretical grounds: indeed
the free $\Lambda$ decay in the {\it Standard Model} can occur
through both $\Delta I=1/2$ and $\Delta I=3/2$
transitions, with comparable strengths. Moreover, the effective
4-quark weak interaction derived from the {\it Standard Model} including
perturbative QCD corrections gives too small $A_{1/2}/A_{3/2}$
ratios ($\simeq 3\div 4$, as calculated at the hadronic scale of about
$1$ GeV by using renormalization group techniques). One can guess that
non--perturbative QCD effects at low energy  (such as hadron structure
and reaction mechanism) and/or final state interactions could be
responsible for the enhancement of the $\Delta I=1/2$ amplitude and/or
for the suppression of the $\Delta I=3/2$ one. Unfortunately these low
energy effects are more difficult to handle: for example chiral
perturbation theory, which is frequently employed for describing hadronic
phenomena in the low energy regime, 
even if used together with perturbative QCD corrections, is
not able to reproduce the hyperon non--leptonic weak decay rates.

Let us turn now to the weak decay of hypernuclei: here the mesonic decay
is disfavoured by the Pauli principle, particularly in heavy systems, as
the momentum of the final nucleon in (\ref{lambdadec})
is only about 100~MeV/c (much smaller than the average Fermi momentum).
Yet, it is the dominant decay channel in $s$-shell hypernuclei. 
From theoretical calculations
and measurements there is evidence \cite{Os98} that in nuclei the ratio
$\Gamma_{\pi^-}/\Gamma_{\pi^0}$  strongly oscillates around the value 2
predicted by the $\Delta I=1/2$ rule in a symmetric nucleus ($N=Z$).
However, these oscillations are essentially due to nuclear shell
effects and might not be directly related to the weak process itself.

In all but the lightest hypernuclei, the dominant weak decay mode
is the non--mesonic one, induced by the interaction of the hyperon
with one or more nucleons, e.g.
\beq
\label{non-mes1}
\Lambda N\to NN  \hspace{0.3cm} (\Gamma_1) , 
\eeq
\beq
\label{non-mes2}
\Lambda NN\to NNN  \hspace{0.3cm} (\Gamma_2) .
\eeq

In this letter we wish to explore the validity of the $\Delta I=1/2$ rule
in the non--mesonic (one--nucleon induced) $\Lambda$--decay: since this
process can only occur in nuclei, its analysis is complicated by
nuclear structure effects, which can alter the simple balance of the
isospin change based on angular momentum couplings.

The one--nucleon induced mechanism concerns  the following
isospin channels:
\begin{eqnarray}
\Lambda n & \to & nn \hspace{0.3cm} (\Gamma_n) , \\ \nonumber
\Lambda p & \to & np \hspace{0.3cm} (\Gamma_p) ,
\end{eqnarray}
where, in the average,  the final nucleons have large momenta
(about 420 MeV). Although the two--nucleon stimulated decay (which
has not been detected yet) is non--negligible (for $p$--shell to heavy
hypernuclei $\Gamma_2$ is about 15\% of the total rate, 
while for $s$--shell hypernuclei its contribution is $\simeq 5$\%
\cite{Al99,Ga99}), in the following we shall consider the process 
$\Lambda N\to NN$ only. 

Direct measurements of the $\Lambda N\to NN$ process
are very difficult to perform, due to the short lifetime
of the hyperons, which gives flight paths limited to less than 10 cm.
The inverse reaction
$pn\rightarrow p\Lambda$ in free space is now under investigation
at COSY and KEK. Nevertheless, only
the precise measurement of the non--mesonic width 
$\Gamma_{NM}=\Gamma_n+\Gamma_p$ in
$s$-shell hypernuclei is nowadays, at least potentially, a good tool to
study the spin and isospin dependence and the validity of the
$\Delta I=1/2$ rule in the $\Lambda N\to NN$ weak interaction.
In $s$--shell hypernuclei all the nucleons are confined (as the hyperon)
into the $s$--level (hence, only the relative 
orbital angular momentum $L=0$ is present in the $\Lambda N$ initial state), 
while complications arise with increasing mass number, due to the
appearance of more $\Lambda N$ states and of the nucleons' rescattering
inside the nucleus, which complicates the kinematic of the measured nucleons.

Nowadays, the main problem concerning the weak decay of
$\Lambda$--hypernuclei is to reproduce the experimental values for the
ratio ${\Gamma}_n/{\Gamma}_p$ between the neutron-- and the 
proton--induced widths (on the contrary, the total non--mesonic widths are well
explained by the available models \cite{Pa97,Ga99,It98,Ok99,Al99}).
The theoretical calculations underestimate the central data points for all
considered hypernuclei, although the large experimental error bars do
not permit any definitive conclusion.
The data are quite limited and not precise since it is difficult to
detect the products of the non--mesonic decays, especially for the
neutron--induced one.
Moreover, the present experimental energy resolution for the detection of the
outgoing nucleons does not allow to identify the final state of the
residual nucleus in the processes
$^A_{\Lambda}{\rm Z}\rightarrow {^{A-2}{\rm Z}} + nn$ and
$^A_{\Lambda}{\rm Z}\rightarrow {^{A-2}({\rm Z-1})} + np$.
As a consequence, the measurements supply decay rates averaged over
several nuclear final states.

In order to solve the $\Gamma_n/\Gamma_p$ puzzle, many attempts have
been made up to now, but without success. Among these we recall the
introduction of mesons heavier than the pion
in the ${\Lambda}N\rightarrow NN$ transition potential
\cite{Sh94,Du96,Pa97,It98},
the inclusion of interaction terms that explicitly 
violate the ${\Delta}I=1/2$ rule \cite{Ma95,Pa98} and the
description of the
short range baryon--baryon interaction in terms of quark degrees of freedom
(by using a hybrid quark model in \cite{Ch83} and a direct quark
mechanism in \cite{Ok99}),
which automatically introduce $\Delta I=3/2$ contributions.

As we shall see in the following, the analysis of the non--mesonic
decays in $s$--shell hypernuclei is important both
for the solution of the $\Gamma_n/\Gamma_p$ 
puzzle and for testing the validity of the
related $\Delta I=1/2$ rule.
Let us start by considering the possible $\Lambda N\rightarrow NN$
transitions  in $s$--shell hypernuclei: since  the $\Lambda N$ pair
is in a state with relative orbital angular momentum $L=0$, the only
possibilities are the following ones 
(we use the spectroscopic notation $^{2S+1}L_J$):
\begin{eqnarray}
\label{partial}
^1S_0 &\rightarrow &^1S_0 \hspace{0.3cm}(I_f=1) \\
      &\rightarrow &^3P_0 \hspace{0.3cm}(I_f=1) \nonumber \\
^3S_1 &\rightarrow &^3S_1 \hspace{0.3cm}(I_f=0) \nonumber \\
      &\rightarrow &^1P_1 \hspace{0.3cm}(I_f=0) \nonumber \\
      &\rightarrow &^3P_1 \hspace{0.3cm}(I_f=1) \nonumber \\
      &\rightarrow &^3D_1 \hspace{0.3cm}(I_f=0) . \nonumber
\end{eqnarray}
The $\Lambda n\rightarrow nn$ process has final states with isospin 
$I_f=1$ only, while for $\Lambda p\rightarrow np$ both $I_f=1$ and
$I_f=0$ are allowed.

Classically, the interaction probability of a particle which crosses an
infinite homogeneous system of thickness $ds$ is $dP=ds/\lambda$,
where $\lambda=1/(\sigma \rho)$ is the mean free path of the particle,
$\sigma$ is the relevant cross section and $\rho$ is the density of
the system.
The width $\Gamma_{NM}=dP/dt$ for the process $\Lambda N\to NN$,
is then given by:
\beq
\label{classic}
\Gamma_{NM}=v\sigma \rho ,
\eeq
$v$ being the $\Lambda$ velocity in the rest frame of the homogeneous system. 
For a finite nucleus, one can weight its density $\rho(\vec r)$,
within the semi-classical approximation, 
by the  $\Lambda$ wave function in the hypernucleus,
$\psi_{\Lambda}({\vec r})$:
\beq
\Gamma_{NM}=\langle v\sigma \rangle \int d{\vec r} \rho({\vec r}) \mid
\psi_{\Lambda}({\vec r})\mid^2 ,
\eeq
where $\langle\, \rangle$ denotes an average over spin and isospin
states. In the above equation the nuclear density is normalized to the
mass number $A=N+Z$, hence the integral gives
the average nucleon density $\rho_{A+1}$ at the position of the $\Lambda$ 
particle. In this scheme, the non--mesonic width 
of the hypernucleus $^{A+1}_{\Lambda}Z$ is then:
\beq
\Gamma_{NM}(^{A+1}_{\Lambda}Z)=
\frac{N\overline{R}_n+Z\overline{R}_p}{A}\rho_{A+1}
\equiv \Gamma_n(^{A+1}_{\Lambda}Z ) + \Gamma_p(^{A+1}_{\Lambda}Z ),
\eeq
where $\overline{R}_n$ ($\overline{R}_p$) denotes the spin--averaged rate
for the neutron--induced (proton--induced) process appropriate for the 
considered hypernucleus. 

Then, by introducing the rates $R_{NJ}$ for spin--singlet ($R_{n0}$, $R_{p0}$) and 
spin--triplet ($R_{n1}$, $R_{p1}$) interactions, the non--mesonic decay widths
of $s$--shell hypernuclei are \cite{Bl62}:
\begin{eqnarray}
\label{phen}
\Gamma_{NM}(^3_{\Lambda}{\rm H})=
\left(3R_{n0}+R_{n1}+3R_{p0}+R_{p1}\right)\frac{\rho_3}{8} ,\\ 
\Gamma_{NM}(^4_{\Lambda}{\rm H})=
\left(R_{n0}+3R_{n1}+2R_{p0}\right)\frac{\rho_4}{6} ,\nonumber \\ 
\Gamma_{NM}(^4_{\Lambda}{\rm He})=
\left(2R_{n0}+R_{p0}+3R_{p1}\right)\frac{\rho_4}{6} ,\nonumber \\
\Gamma_{NM}(^5_{\Lambda}{\rm He})=
\left(R_{n0}+3R_{n1}+R_{p0}+3R_{p1}\right)\frac{\rho_5}{8} , \nonumber
\end{eqnarray}
where we have taken into account that the hypernuclear total angular
momentum is 0 for $^4_{\Lambda}$H and $^4_{\Lambda}$He and 1/2 for
$^3_{\Lambda}$H and $^5_{\Lambda}$He.
In terms of the rates associated to the partial--wave
transitions (\ref{partial}), the $R_{NJ}$'s of eq.~(\ref{phen}) are:
\begin{eqnarray}
R_{n0}&=&R_n(^1S_0)+R_n(^3P_0) , \\
R_{p0}&=&R_p(^1S_0)+R_p(^3P_0) , \nonumber \\ 
R_{n1}&=&R_n(^3P_1) , \nonumber \\
R_{p1}&=&R_p(^3S_1)+R_p(^1P_1)+R_p(^3P_1)+R_p(^3D_1) , \nonumber 
\end{eqnarray}
the quantum numbers of the $NN$ final state being reported in brackets.

If we assume that the $\Lambda N\to NN$ weak interaction occurs with a change
$\Delta I=1/2$ of the isospin, the following relations 
(simply derived by angular momentum coupling coefficients)
hold among the rates for transitions to isospin 1 final states:
\beq
\label{uno1}
R_n(^1S_0)=2R_p(^1S_0), \hspace{0.2cm} R_n(^3P_0)=2R_p(^3P_0) ,
\hspace{0.2cm} R_n(^3P_1)=2R_p(^3P_1) ,
\eeq
then:
\beq
\label{uno2}
\frac{R_{n1}}{R_{p1}}\leq\frac{R_{n0}}{R_{p0}}=2 .
\eeq
For pure $\Delta I=3/2$ transitions, the factor 2 in the right hand
side of eqs.~(\ref{uno1}), (\ref{uno2}) should be replaced by 1/2.

Let us now introduce the ratios:
\beq
r=\frac{\langle I_f=1||A_{1/2}||I_i=1/2\rangle}
{\langle I_f=1||A_{3/2}||I_i=1/2\rangle} 
\eeq
between the $\Delta I=1/2$ and $\Delta I=3/2$ $\Lambda N\to NN$
transition amplitudes for isospin 1 final states
($r$ being real, as required by time reversal invariance) 
and:
\beq
\label{lam}
\lambda=\frac{\langle I_f=0||A_{1/2}||I_i=1/2\rangle}
{\langle I_f=1||A_{3/2}||I_i=1/2\rangle} .
\eeq
Then, for a general $\Delta I=1/2$--$\Delta I=3/2$ mixture, we have:
\beq
\label{mixt1}
\frac{R_{n1}}{R_{p1}}= \frac{4r^2-4r+1}{2r^2+4r+2+6\lambda^2}\leq
\frac{R_{n0}}{R_{p0}}=\frac{4r^2-4r+1}{2r^2+4r+2} \, .
\eeq
By using equations (\ref{phen}) and (\ref{mixt1}) 
together with the available experimental data, it is possible to extract the 
spin and isospin behavior of the $\Lambda N\to NN$ interaction without
resorting  to a detailed knowledge of the interaction mechanism.
It is worth pointing out that, according to the above relations, the
value of $\Gamma_n/\Gamma_p$ is not a universal function of the ratios
$R_{nJ}/R_{pJ}$: it rather depends upon the spin, isospin and structure
of the considered hypernucleus. A ratio   $\Gamma_n/\Gamma_p <2$ does not
necessarily imply that the $\Delta I=1/2$ rule is valid; on the contrary
the presence of $\Delta I=3/2$ transitions can even lower this ratio.
The opposite situation, $\Gamma_n/\Gamma_p > 2$, cannot be simply explained
through a violation of the  $\Delta I=1/2$ rule, and more complicated
reaction mechanisms, including the two--nucleon induced decay and the
nucleons final state interactions, are likely to play a role.

We must notice that this analysis makes use of several assumptions.
For example, the decay is treated incoherently
on the stimulating nucleons, within a simple 4--barions point interaction
model, and the nucleon final state interactions are neglected.
Moreover, the calculation requires the knowledge of the
nuclear density at the hyperon position (here, in particular, the same
density is employed for $^4_{\Lambda}$H and $^4_{\Lambda}$He).
Since the available data have large error bars, the above approximations
can be considered as satisfactory.
This phenomenological model was introduced by Block and Dalitz \cite{Bl62}.
More recently, the analysis has been updated by other authors
\cite{Do87,Co90,Sc92}. These works suggested a sizeable violation of the
$\Delta I=1/2$ rule, although the authors pointed out the need for
more precise data to draw definitive conclusions.

\begin{table}[t]
\begin{center}
\caption{Experimental data (in units of $\Gamma^{\rm free}_{\Lambda}$) 
for $s$-shell hypernuclei.}
\label{data}
\begin{tabular}{c|c c c c c}
\mc {1}{c|}{} &
\mc {1}{c}{$\Gamma_n$} &
\mc {1}{c}{$\Gamma_p$} &
\mc {1}{c}{$\Gamma_{NM}$} &
\mc {1}{c}{$\Gamma_n/\Gamma_p$} & 
\mc {1}{c}{Ref.} \\ \hline\hline
$^4_{\Lambda}$H    &       &     & $0.22\pm 0.09$ & & reference value (average)\\
                   &       &     & $0.17\pm 0.11$ &  &  KEK \cite{Ou98}\\
                   &       &     & $0.29\pm 0.14$ &  & \cite{Bl62}\\ \hline
$^4_{\Lambda}$He   &$0.04\pm 0.02$   & $0.16\pm 0.02$ & $0.20\pm 0.03$
& $0.25\pm 0.13$   &  BNL \cite{Ze98} \\ \hline
$^5_{\Lambda}$He     &$0.20\pm 0.11$  & $0.21\pm 0.07$ & $0.41\pm 0.14$ &
$0.93\pm 0.55$    & BNL \cite{Sz91} \\ 
\end{tabular}  
\end{center}
\end{table}

Here we shall use more recent data (which are summarized in table~\ref{data})
and we shall employ a different analysis.
Unfortunately, no data are available on the non--mesonic decay of
hyper-triton. We use the BNL data \cite{Ze98,Sz91} for $^4_{\Lambda}$He
and $^5_{\Lambda}$He together with the {\it reference value}
of table~\ref{data} for $^4_{\Lambda}$H.
This last number is the average of the previous estimates 
of refs.~\cite{Ou98,Bl62}, which have not been obtained
from direct measurements but rather by using theoretical constraints.

We have then 5 independent data which allow to fix, from eq.~(\ref{phen}),
the 4 rates $R_{N,J}$ and $\rho_4$. Instead, the density $\rho_5$ also
entering into eq.~(\ref{phen}), has been {\it estimated} to be 
$\rho_5=0.045$ fm$^{-3}$ by employing the $\Lambda$ wave function of
ref.~\cite{St93} (obtained through a quark model description of the
$\Lambda N$ interaction) and the Gaussian density for $^4$He that 
reproduces the experimental mean square radius of the nucleus.
For $^4_{\Lambda}$H and $^4_{\Lambda}$He no realistic hyperon wave
function is available and we can obtain the value of $\rho_4$
from the data of table~\ref{data}, by imposing that [see eq.~(\ref{phen})]:
\beq
\frac{\Gamma_p(^5_{\Lambda}{\rm He})}{\Gamma_p(^4_{\Lambda}{\rm He})}=
\frac{3}{4} \frac{\rho_5}{\rho_4} .
\eeq
This yields $\rho_4=0.026$ fm$^{-3}$.  Furthermore, the best determination
of the rates $R_{N,J}$ is obtained from the relations for the observables:
\beq
\Gamma_{NM}(^4_{\Lambda}{\rm H}) , \hspace{0.3cm}
\Gamma_{NM}(^4_{\Lambda}{\rm He}) ,
\hspace{0.3cm} \Gamma_{NM}(^5_{\Lambda}{\rm He}) , \hspace{0.3cm} 
\frac{\Gamma_n}{\Gamma_p}(^4_{\Lambda}{\rm He}) ,
\label{wan}
\eeq
which have the smallest experimental uncertainties.
Solving these equations we extracted the following partial rates
(the decay rates of eq.~(\ref{phen}) are considered in units of the
free $\Lambda$ decay width):
\begin{eqnarray}
\label{results}
R_{n0}&=&(4.7\pm 2.1)\:{\rm fm}^3 , \\
R_{p0}&=&(7.9^{+16.5}_{-7.9})\:{\rm fm}^3 , \\ 
R_{n1}&=&(10.3\pm 8.6)\:{\rm fm}^3 , \\ 
\label{results2}
R_{p1}&=&(9.8\pm 5.5)\:{\rm fm}^3 ,  
\end{eqnarray}
The errors have been obtained with the standard formula:
\beq
\delta[O(r_1,..,r_N)]=\sqrt{\sum_{i=1}^{N} 
\left(\frac{\partial O}{\partial r_i}\delta r_i\right)^2} ,
\eeq
namely by treating the data as independent and uncorrelated ones.
Due to the large relative errors 
implied in the extraction of the above rates,
the Gaussian propagation of the uncertainties has to be regarded as a 
poor approximation. 

For the ratios of eq.~(\ref{mixt1}) we have then:
\beq
\label{gen1}
\frac{R_{n0}}{R_{p0}}=0.6^{+1.3}_{-0.6} ,
\eeq
\beq
\label{gen2}
\frac{R_{n1}}{R_{p1}}=1.0^{+1.1}_{-1.0} .
\eeq
while the ratios of the spin--triplet to the spin--singlet interaction
rates are:
\beq
\label{gen3}
\frac{R_{n1}}{R_{n0}}=2.2\pm 2.1 ,
\eeq
\beq
\label{gen4}
\frac{R_{p1}}{R_{p0}}=1.2^{+2.7}_{-1.2} .
\eeq
The large uncertainties do not allow
to drawn definite conclusions about the possible violation of the
$\Delta I=1/2$ rule and the spin--dependence of the transition rates.
Eqs.~(\ref{gen1}) and (\ref{gen2}) are still compatible
with eq.~(\ref{uno2}), namely with the $\Delta I=1/2$ rule, although
the central value in eq.~({\ref{gen1}}) is more in
agreement with a pure $\Delta I=3/2$ transition ($r\simeq 0$) or 
with $r\simeq 2$ [see eq.~(\ref{mixt1})].
Actually, eq.~(\ref{gen1}) is compatible with $r$ 
in the range $-1/4\div 40$, while the ratio $\lambda$ of 
eq.~(\ref{mixt1})  is completely undetermined.

By using the results of eqs.~(\ref{results})-(\ref{results2}) into
eq.~(\ref{phen}) we can predict the neutron to proton ratio for
hyper-triton, which turns out to be:
\beq
\frac{\Gamma_n}{\Gamma_p}(^3_{\Lambda}{\rm H})=0.7^{+1.1}_{-0.7} ,
\eeq
and, by using $\rho_3=0.001$ fm$^{-3}$ \cite{Bl62},
\beq
\Gamma_{NM}(^3_{\Lambda}{\rm H})=0.007\pm 0.006 .
\eeq
The latter is of the same order of magnitude of the
detailed 3-body calculation of ref.~\cite{Go97},
which provides a non-mesonic width equal to 
1.7\% of the free $\Lambda$ width.

The compatibility of the data with the $\Delta I=1/2$ rule can be
exploited in a different way.
By {\it assuming} this rule, $R_{n0}/R_{p0}=2$; then one can use the
three observables (instead of the four in eq.~(\ref{wan})):
\beq
\Gamma_{NM}(^4_{\Lambda}{\rm He}) ,
\hspace{0.3cm} \Gamma_{NM}(^5_{\Lambda}{\rm He}) , \hspace{0.3cm}
\frac{\Gamma_n}{\Gamma_p}(^4_{\Lambda}{\rm He}) ,
\eeq
to extract the following partial rates:
\begin{eqnarray}
R_{n0}&=&(4.7\pm 2.1)\:{\rm fm}^3 , \\
R_{p0}\equiv R_{n0}/2&=&(2.3\pm 1.0)\:{\rm fm}^3 , \\
R_{n1}&=&(10.3\pm 8.6)\:{\rm fm}^3 , \\
R_{p1}&=&(11.7\pm 2.4)\:{\rm fm}^3 .
\end{eqnarray}
These values are compatible with the ones in 
eqs.~(\ref{results})-(\ref{results2}) (actually $R_{n0}$ and $R_{n1}$
are unchanged with respect to the above derivation).
For pure $\Delta I=1/2$ transitions the spin--triplet interactions
seem to dominate over the spin--singlet ones:
\beq
\frac{R_{n1}}{R_{n0}}=2.2\pm 2.1 ,
\eeq
\beq
\frac{R_{p1}}{R_{p0}}=5.0\pm 2.4 .
\eeq
Moreover, since:
\beq
\frac{R_{n1}}{R_{p1}}=0.9\pm 0.8 ,
\eeq
from eq.~(\ref{mixt1}) one obtains the following estimate
for the ratio between the $\Delta I=1/2$ amplitudes:
\beq
\left|\frac{\lambda}{r}\right|\equiv
\left| \frac{\langle I_f=0||A_{1/2}||I_i=1/2\rangle}   
{\langle I_f=1||A_{1/2}||I_i=1/2\rangle}\right| \simeq \frac{1}{3.7}\div 2.3 .
\eeq
Finally, the other independent observable (which has not been utilized here)
is predicted to be:
\beq
\Gamma_{NM}(^4_{\Lambda}{\rm H})=0.17\pm 0.11 ,
\eeq
in good agreement with the experimental data of table~\ref{data},
within $\simeq 0.6\, \sigma$ deviation. This implies that the data are 
consistent with
the $\Delta I=1/2$ rule at the level of 60\%. Or, in other words,
the $\Delta I=1/2$ rule is excluded at the 40\% confidence
level.

The phenomenological model of Block and Dalitz can be extended
to hypernuclei beyond the $s$--shell; for the sake of illustration
we consider here $^{12}_{\Lambda}$C. In table~\ref{data2} the data on
the non--mesonic decay of this hypernucleus are quoted.
\begin{table}[t]
\begin{center}
\caption{Experimental data for $^{12}_{\Lambda}$C.}
\label{data2}
\begin{tabular}{c c c}
\mc {1}{c}{$\Gamma_{NM}$} &
\mc {1}{c}{$\Gamma_n/\Gamma_p$} &
\mc {1}{c}{Ref.} \\ \hline\hline
$1.14\pm 0.20$ & $1.33^{+1.12}_{-0.81}$ & BNL \cite{Sz91} \\
$0.89\pm 0.18$ & $1.87^{+0.67}_{-1.16}$ & KEK \cite{No95}  \\
$1.01\pm 0.13$ & $1.61^{+0.57}_{-0.66}$ & average \\ 
\end{tabular}
\end{center}
\end{table}
Within the present framework, the relevant decay rate can be written
in the following form:
\beq
\label{carbonio}
\Gamma_{NM}(^{12}_{\Lambda}{\rm C})=\frac{\rho^s_{12}}{\rho_5}
\Gamma_{NM}(^5_{\Lambda}{\rm He})+
\frac{\rho^p_{12}}{7}\left
[3\overline{R}_n(P)+4\overline{R}_p(P)\right] ,
\eeq
where $\rho^s_{12}$ ($\rho^p_{12}$) is the average $s$--shell 
($p$--shell) nucleon density at the hyperon position, while
$\overline{R}_n(P)$ [$\overline{R}_p(P)$] is the spin--averaged
$P$--wave neutron--induced (proton--induced) rate. By using the 
previous results from $s$--shell hypernuclei
and the average values in tab.~\ref{data2}, we obtain:
\begin{eqnarray}
\overline{R}_n(P)&=&(18.3\pm 10.7)\:{\rm fm}^3 , \\
\label{errecarb}
\overline{R}_p(P)&=&(3.6^{+12.6}_{-3.6})\:{\rm fm}^3 .
\end{eqnarray}
The densities $\rho^s_{12}$ ($=0.064$ fm$^{-3}$) and 
$\rho^p_{12}$ ($=0.043$ fm$^{-3}$) have been calculated from the
appropriate $s$-- and $p$--shell Woods--Saxon nucleon wave functions.
The $s$-- and $p$--shell contributions in eq.~(\ref{carbonio})
are $0.58\pm 0.20$ and $0.43\pm 0.24$, respectively.
The former is calculated starting
from the experimental non--mesonic decay width
for $^5_{\Lambda}$He, the latter by subtracting the $s$--shell 
contribution from the average value of table~\ref{data2} 
for $\Gamma_{NM}$. The central values quoted above
are in disagreement with 
the detailed calculation of refs.~\cite{Pa97,It98},
where the contribution of the $P$ partial wave to $\Gamma_{NM}$
is estimated to be only $5\div 10$\% in $p$--shell hypernuclei.
However, due to the large uncertainties,
at the $2 \sigma$ level our result is compatible with a negligible
$P$--wave contribution. 
Moreover, we must notice that in eq.~(\ref{carbonio}) the contribution
of $S$--wave $\Lambda N$ relative states originating from the interaction
with $p$--shell nucleons is neglected.
It is possible to include this contribution by changing
$\rho^s_{12}\to \rho^s_{12}+\alpha \rho^p_{12}$ in the first term in the
right hand side of eq.~(\ref{carbonio}), $\alpha$ being the  fraction
of $1p$ nucleons which interact with the $\Lambda$ in $S$ relative wave.
In order to reproduce a 10\% contribution of the $P$-wave interaction to
$\Gamma_{NM}(^{12}_{\Lambda}{\rm C})$
[second term in the right hand side of eq.~(\ref{carbonio})], 
a large $\alpha$ is required: $\alpha\simeq 0.8$. 

Other applications can be considered in heavier hypernuclei, providing
one neglects $\Lambda N$ interactions in $D, F,$ etc waves; then, by
using the description and results of eqs.~(\ref{carbonio})-(\ref{errecarb}),
we can easily predict, e.g., the non--mesonic rate for $^{56}_{\Lambda}$Fe,
with the result:
\beq
\label{iron}
\Gamma_{NM}(^{56}_{\Lambda}{\rm Fe})=\frac{\rho^s_{56}}{\rho_5}
\Gamma_{NM}(^5_{\Lambda}{\rm He})+\frac{\rho^p_{56}}{2}
\left[\overline{R}_n(P)+\overline{R}_p(P)\right]=1.48^{+0.59}_{-0.45} ,
\eeq
where $\rho^s_{56}=0.087$ fm$^{-3}$ ($\rho^p_{56}=0.063$ fm$^{-3}$) now  
embodies the contributions from both the $1s$ and $2s$ ($1p$ and $2p$) nucleon
levels. The central value of eq.~(\ref{iron}) 
overestimates the KEK result \cite{Bh98}, 
which measured a total width $\Gamma_T=1.22\pm 0.08$
(note that for iron the mesonic rate is negligible at this level
of accuracy), but the two numbers are
compatible at the $1\sigma$ level. Notice that, should we use 
the prescription (suggested above for $^{12}$C)
$\rho^s_{56}\to \rho^s_{56}+0.8\rho^p_{56}$,
eq.~(\ref{iron}) would yield about the same result for 
$\Gamma_{NM}(^{56}_{\Lambda}{\rm Fe})$ (1.39), but with
a different balance of the $S$-- and $P$--wave terms
and a smaller $\Gamma_n/\Gamma_p$ ratio (1.09 instead of 1.85).

In concluding this letter we wish to stress that the phenomenological
model of Block and Dalitz employed here, in spite of its relative
simplicity, allows to set rather precise constraints on the partial
contributions to the non--mesonic decay width of $s$--shell hypernuclei.
From the available experimental data we have found some limits on the
ratio between $\Delta I=1/2$ and $\Delta I=3/2$ transition amplitudes.
A violation of the $\Delta I=1/2$ rule in the one--nucleon
induced non--mesonic decay  seems to be present at 40\%  confidence level.
Unfortunately the experimental uncertainties are still too large to
allow any definitive conclusion. Similar considerations hold valid
in heavier systems, in which however the present analysis requires
more severe approximations and must be applied with some caution. 

\vspace{1.5cm}
We acknowledge financial support from the MURST.
This work was supported in part by the EEC through TMR Contract CEE-0169.


\vfill\eject

\end{document}